\documentclass[a4paper,11pt]{article}
\usepackage[T2A]{fontenc}
\usepackage[cp1251]{inputenc}
\usepackage{amsmath,amsthm,amssymb}
\usepackage{graphics}
\usepackage{graphicx}
\begin{document}
\title{Spin-dependent conductivity of iron-based superconductors in a magnetic field}
\author{M.O. Dzyuba$^{a}$, Yu.N. Chiang$^{a}$, D.A. Chareev$^{b}$, A.N. Vasiliev$^{c}$\\
\emph{$^{a}$ B. I. Verkin Institute for Low Temperature Physics and Engineering, }\\
\emph{National Academy of Sciences of Ukraine,}\\
\emph{47 Lenin ave., Kharkov 61103, Ukraine}\\
\emph{$^{b}$Institute of Experimental Mineralogy RAS,}\\
\emph{Chernogolovka, 142432, Russian Federation}\\
\emph{$^{c}$Lomonosov Moscow State University,}\\
\emph{GSP-1, Leninskie Gorky, Moscow 119991, Russian Federation}\\
\emph{E-mail:dzyuba@ilt.kharkov.ua}}
\date{}
\maketitle
\begin{abstract}
We report the results of a study of magnetic field features of electron transport in heterojunctions with NS
boundary inside iron-based superconductors, represented by a binary phase of $\alpha$ - FeSe and oxyarsenide
pnictide LaO(F)FeAs. We used the ability of self magnetic field of the transport current to partially destroy
superconductivity, no matter how low the field may be, in the NS interface area, where, due to the proximity
effect, the superconducting order parameter, $\Delta$, disperses from 1 to 0 within the scale of the
Ginzburg-Landau coherence length.

The following features of transport were found: (i) at $T<T_{c}$, magnetoresistance in systems with different
superconductors has different sign; (ii) sign and magnitude of the magnetoresistance depend on the magnitude of
current and temperature, and (iii) in all operating modes where the contribution from Andreev reflection is
suppressed ($(T + eV) \gtrsim \Delta$), the hysteresis of the magnetoresistance is present.

Based on the results of the experiment and analysis it has been concluded that there is a long-range magnetic
order in the ground normal state of the iron-based superconductors studied, in the presence of itinerant
magnetism of conduction electrons which determines the possibility of anisotropic spin-dependent exchange
interaction with the local magnetic moments of the ions.
\end{abstract}

Search and study of low-symmetry systems with crystal symmetry of "layered" type, characterized by anisotropic
electrical and magnetic properties leading to superconductivity, is one of the most urgent problems of condensed
matter physics. Currently, superconductivity is found in the vast family of such compounds containing a wide
range of rare earths, pnictogens, chalcogens, and transition elements Mn, Fe, Co, Ni, Cu, Ru. In particular,
observation of superconductivity in iron-based structures suggests that it is the anisotropy of the properties
that seems to be an essential condition under which in the same material may coexist magnetic interactions and
interactions that determine the superconducting pairing of excitations in the electronic subsystem.

While the ideas of the crystal structures and nature of coupling in such superconductors are sufficiently
developed and experimentally proved, their magnetic and electronic structures, and especially the nature of their
interaction in the ground state of iron-based superconductors, are still the subject of intense debate (see, eg,
[1, 2]). Therefore, it becomes important to study transport phenomena in the systems containing such
superconductors, as the nature of those phenomena may directly depend on itinerant magnetism of conduction
electrons, capable of interacting with the sublattice of localized magnetic moments of the magnetic atoms.

Here we report the results of studying hysteresis features of electron transport in a magnetic field in the
normal ground state of iron-based superconductors, namely, monocrystalline binary phase of $\alpha$ - FeSe and
oxyarsenide pnictide LaO(F)FeAs in granular form. Both crystal structures share common structural unit of the
same symmetry type, PbO (P4/\emph {nmm}), predetermining the affinity of the exchange interactions in
quasi-two-dimensional electronic bands with nesting areas of Fermi-surface [3 - 5].

\begin{figure}[htb]
  \begin{center}
  \includegraphics[width=10cm]{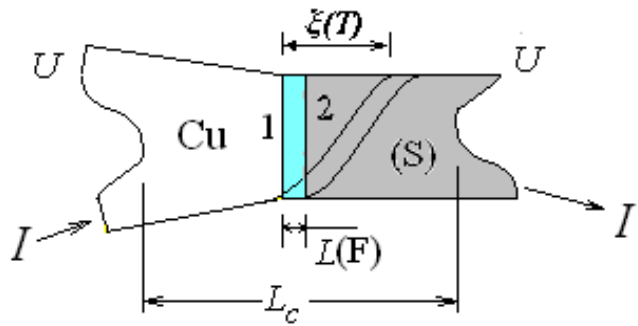}\\
  \caption{Schematic representation of a typical microheterocontact with its distinctive
spatial parameters under the proximity effect. $L_{c} = 1.5 \div 3$ mm is a total length of the contact
corresponding to the distance between the measuring potential probes in the NN state; 1, 2 denote the NS boundary
position without (1) and with (2) a transport current; \emph{L}(F) is a part of the superconductor (S) passed
into the normal state as a result of shift of NS boundary.} \label{1}
\end{center}
\end{figure}
The raw materials of superconductors used to prepare the samples had different structure according to the
technology for their manufacture. Pnictide La[O$_{0.85}$F$_{\sim 0.1}$]FeAs was prepared by solid-phase
synthesis, such as that described in [6], and had a polycrystalline structure of granular type. Iron
monochalcogenide FeSe was made in the single crystal form [7].

As is known, magnetic current-voltage characteristics of the macroscopic granular samples of high-temperature
superconductors exhibit hysteresis, the nature of which is usually associated with a variety of possible
scenarios of current flow, including percolation, tunneling (intergranular or intragranular interfacial), and
intragranular conduction mechanisms special for a given superconductor [8 - 10]. Since typical size of grains
(granules) in materials prepared by a variety of technologies may amount to $d \sim 10^{-4}$ cm, it is clear that
to avoid the first two uncharacteristic scenarios of conductivity, the sample size is desirable to be of the same
order. This condition is satisfied for the point contacts for which, as it has previously been shown by us [11],
the length of the measuring area in NS operating mode is typically of the order of a few microns. This is a
typical mesoscopic scale, non-ballistic due to even smaller elastic mean free paths (details ibid). In this
paper, we present the results of studying the conductance of point-contact samples with ohmic characteristics
similar to those used in [12].

\begin{figure}[htb]
  \begin{center}
  \includegraphics[width=10cm]{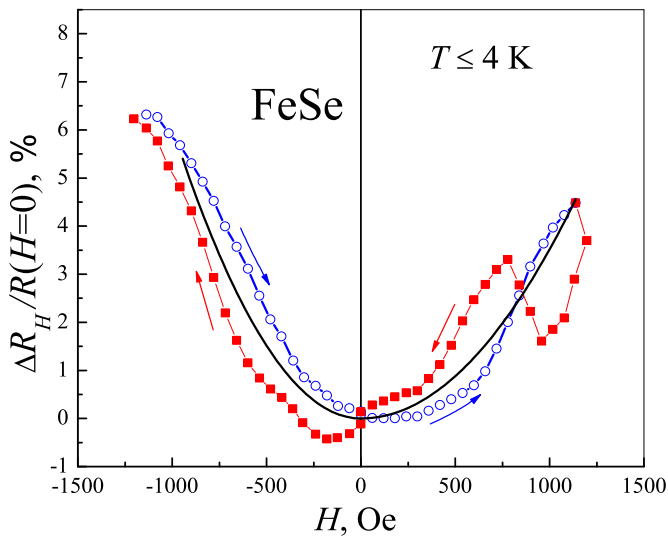}\\
  \caption{Magnetoresistance with hysteresis of the heterojunction Cu/FeSe normalized to the total contact
resistance, with \emph {T} $\lesssim 4$ K and \emph {I} = 1 mA ~ $((T + eV^{*}) \sim \Delta$). Solid curve
represents spin-dependent magnetoresistance in accordance with Eq. (5).} \label{2}
\end{center}
\end{figure}

To convert the superconductor within the heterojunction area into the normal phase, we used the ability of self
magnetic field of the transport current to partially destroy superconductivity, no matter how small the field may
be, in the NS interface area. In it, due to the proximity effect, superconducting order parameter disperses from
1 to 0 within the scale of the Ginzburg-Landau coherence length $\xi_{T} \sim (1-T/T_{c})^{-1/2} \xi_{0}$. This
allows us to shift the dispersion region of the order parameter deep into the superconductor, yielding
defect-free NS boundary inside the superconductor between its normal and superconducting parts (see Fig. 1). In
this, a reasonable scale of spatial dispersion effect comparable with the Ginzburg-Landau coherence length can be
regulated with the help of temperature.

Thus, one can investigate the normal phase of the superconductors in binary NS contacts in the presence of an
ideal N(F)/S boundary (symbol F indicates that a superconductor in its normal phase may be in the magnetic
state), located inside the superconductor. We used this fact in [12] studying the Andreev reflection. In
particular, our results showed that the Ginzburg-Landau parameter for considered superconductors $\kappa =
\lambda_{T}/ \xi_{T}$ was $\geq 1$, and hence, the discussed superconductors must have the properties of type II
superconductors with the London penetration depth $\lambda_{\rm L} \simeq 0.2\ \mu$ [ 13].

From the data of Ref. [12] it also follows that the energy \emph {W} of self magnetic field of measuring current
1 mA in the contact Cu/FeSe, with the cross section $\mathcal {A} \geq 10^{-4} ~ {\rm cm}^{2}$, is of order of or
slightly larger than the energy gap $\Delta$ (in the BCS approximation) for FeSe ($W \geq 0.5\ {\rm meV};\ \Delta
\sim 0.5\ {\rm meV}$). At the same time, in the contact Cu/LaO(F)FeAs of the same geometry, $W \leq \Delta ~
(\Delta_{\rm LaO(F)FeAs} \approx 6 \Delta_{\rm FeSe}$). The normal layer thickness of the superconductor, \emph
{L}(F), formed due to shifting the NS boundary as a result of suppressing superconductivity by self magnetic
field of the transport current $I = 1$ mA appeared to be of the order of $\lambda_{T}$. Therefore, at helium
temperatures, \emph {L}(F) approximately amounts to $(10^{-5} \div 10^{-6}) ~ {\rm cm}$ (at $I=100$ mA, it is ln100
times more [12]) while the elastic mean free path of electrons in contacted materials $l_{el}(5\ K) \sim
10^{-6} \div 10^{-8}$ cm (from the measurements on macroscopic samples).

Thus, presence of a normal region of the superconductor in contact allows one to explore the behavior of electron
transport in the N state of the superconductor including upon application of a magnetic field. To avoid
additional complications associated with possible radical rearrangement of the electron spectrum in strong
magnetic fields, we studied the conductivity of iron-based superconductors in low fields.

\begin{figure}[htb]
  \begin{center}
  \includegraphics[width=10cm]{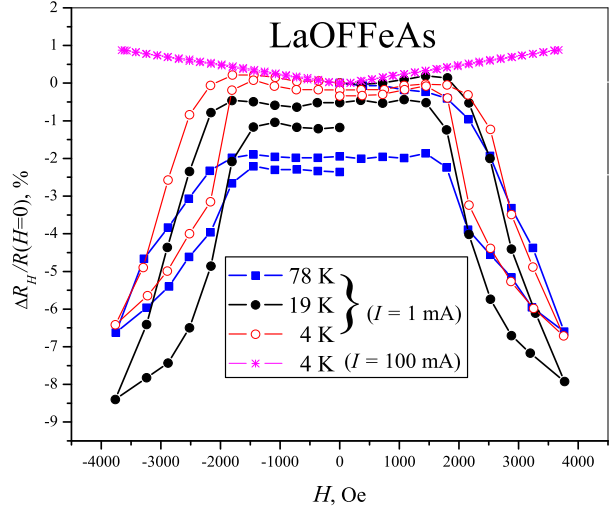}\\
  \caption{Magnetoresistance with hysteresis of heterojunction Cu/LaO(F)FeAs normalized to the total contact resistance. 1)
$\circ$ and $\bullet$ measured at \emph {T} = 4 K and \emph {I} = 1 mA ~ $((T + eV^{*}) < \Delta$); 2) $\ast$ at
\emph {T} = 4 K and \emph {I} = 100 mA ~ $((T + eV^{*}) \sim \Delta$); 3) $\blacksquare$  at \emph {T} = 78 K and
\emph {I} = 1 mA ~ $((T + eV^{*})> \Delta$).} \label{3}
\end{center}
\end{figure}

Figs. 2 and 3 show the behavior of the resistance of heterocontacts Cu/FeSe and Cu/LaO(F)FeAs in low magnetic
fields, $R_{H}$, normalized to the contact resistance in zero magnetic field, $R(H=0)$
, at temperatures \emph {T} below and [for Cu/LaO(F)FeAs] above the superconducting transition temperature
$T_{c}$. For FeSe, $T_{c} \approx 5$ K; for LaO(F)FeAs, $\approx 26$ K. The features we found are as follows.
First, at $T<T_{c}$, magnetoresistance reveals different signs in contacts with different superconductors: in
Cu/FeSe, the sign is positive while in Cu/LaO(F)FeAs it is negative. Second, the magnetoresistance of the same
system, Cu/LaO(F)FeAs, has different signs at different currents: at higher current 100 mA, the addition to the
resistance, $\delta R_{H}$, is positive even at $T<T_{c}$ (see Fig. 3, asterisks) while at lower current 1 mA it
is negative at temperatures \emph {T} both below and above $T_{c}$. In heterocontact Cu/FeSe, the same addition
is predominantly positive in the whole range of currents and in comparable fields, though there exists some
interval of fields wherein the additive is negative. Third, in both systems magnetoresistance exhibits
hysteresis. On the example of heterojunction Cu/LaO(F)FeAs, the first two features in the behavior of
magnetoresistance, for greater clarity, reproduced in Fig. 4 without hysteresis and only for one of the field
direction.

 \begin{figure}[htb]
  \begin{center}
  \includegraphics[width=10cm]{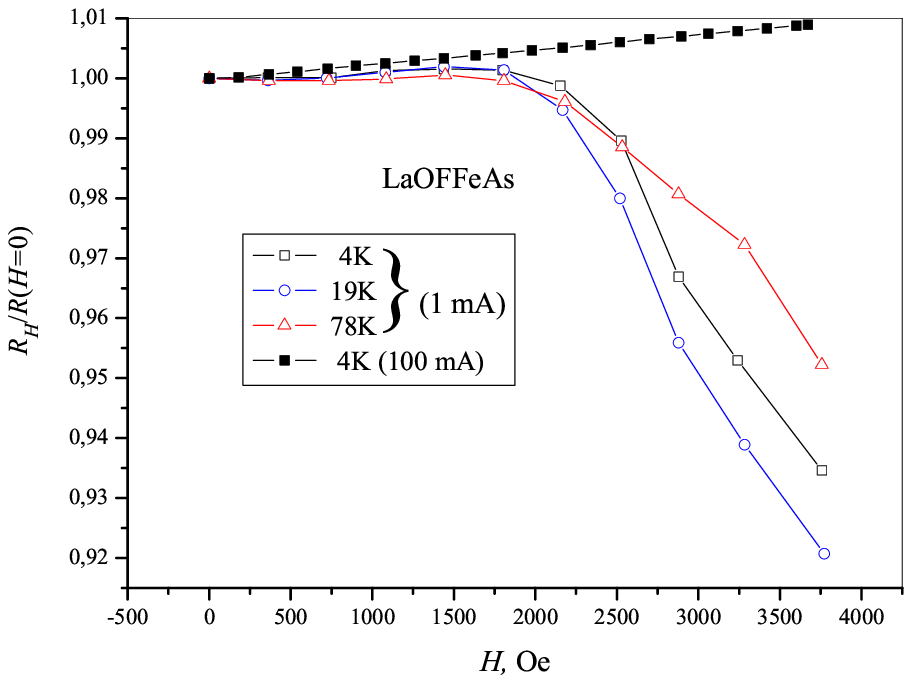}\\
  \caption{Averaged resistance in a magnetic field for heterojunction Cu/LaO(F)FeAs normalized to the total
contact resistance. 1) $\circ$ and $\Box$ correspond \emph {T} = 4 K and \emph {I} = 1 mA ~ $((T +
eV^{*})<\Delta$); 2) $\blacksquare$ \emph {T} = 4 K and \emph {I} = 100 mA ~ $((T + eV^{*}) \sim \Delta$); 3)
$\vartriangle$ \emph {T} = 78 K and \emph {I} = 1 mA ~ $((T + eV^{*})> \Delta$).} \label{4}
\end{center}
\end{figure}

As noted above, effective (measured) length of investigated contacts does not exceed a few microns. In NS mode it
includes four areas: highly conductive normal metal (Cu), oxide barrier, and two border areas of iron-based
superconductor (in the normal and superconducting states). In this structure, only superconductor in the ground
normal state may be responsible for negative magnetoresistance at low fields: in accordance with the
characteristics of other parts of a contact, copper probes and oxide barriers [12], their magnetoresistance can
reveal only slightly increasing behavior in weak magnetic fields.

It is known that an increase in conductivity in a magnetic field (except, perhaps, for field values corresponding
to large Zeeman energies, capable of, for example, inducing metamagnetism), leading to negative
magnetoresistance, indicates either the presence of spin-dependent effects in transport of itinerant conduction
electrons [14] or the degradation of weak-localization interference addition to the resistance [15]. Assessing
the weak-localization interference correction to the conductivity, we find that the negative contribution to
magnetoresistance of the contacts under elastic scattering of electrons (at helium temperatures) due to the
destruction of interference self-intersecting trajectories of electrons by a magnetic field in the studied range
of fields can be of the following order of magnitude:

\begin{equation}\label{1}
    \frac{\Delta \rho_{H}}{\rho (H=0)} = -\Delta \sigma_{H}\cdot\rho_{H} \approx -5\cdot (10^{-5}\div
    10^{-3}),
\end{equation}
where $\Delta \sigma_{H} = \sigma_{H} - \sigma (H = 0) \sim (e^{2}/ \hbar) (eH/ \hbar c)^{1/2}; H = 2 \cdot
10^{3}$ Oe, while the possible values of $\rho (H = 0)$ for the materials comprising the contact fall into the
limits $(1 \div 100) ~ \mu \Omega \cdot$cm (from measurements on bulk samples). Comparing this with the data from
Figs. 2 and 3 we see that this contribution compared with the experimental values of negative addition to the
resistance, even being normalized to total resistance of the contacts studied, is insufficient to explain the
observed behavior of the magnetoresistance. Consequently, the most likely cause of the observed negative
magnetoresistance is spin- dependent nature of the transport of conduction electrons under conditions of their
itinerant magnetism. This spatially relates the change in resistance, $\delta R_{H}$, with the contact part
occupied by two neighboring regions of a superconductor - one in the normal state (\emph {L}(F)) and the other in
the superconducting state.

\begin{figure}[htb]
  \begin{center}
  \includegraphics[width=10cm]{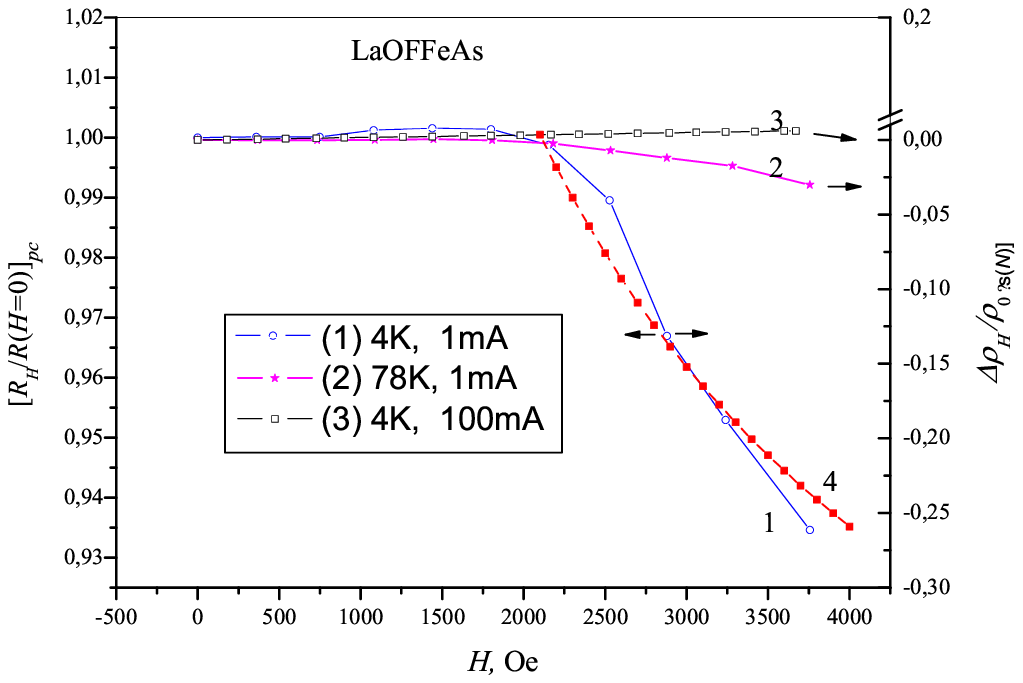}\\
\caption{Averaged magnetoresistance normalized to the total resistance of the contact (left axis) and that
normalized to the resistance of the normal parts of the superconductor (right axis) for heterojunction
Cu/LaO(F)FeAs: 1 - $((T + eV^{*})\ll \Delta$); 2 - $((T + eV^{*})>\Delta$); 3 - $((T + eV^{*}) \sim \Delta$); 4 -
\emph {fit} of the curve 1 according with Eq. (2).} \label{5}
\end{center}
\end{figure}

With this in mind, a correct idea of the scale and behavior of the magnetoresistance $\delta \rho_{H}/ \rho(H =
0)$ we obtain relating a change in resistance $\delta R_{H}$ not to a total contact resistance, as was done in
plotting the curves in Figs. 2, 3, and 4, but only to the resistance of normal areas of a superconductor in
contact: in NS regime (at $T<T_{c}$) - to the resistance $R_{L({\rm F})}\ (H = 0)$ of the \emph {N} (F) area,
$L({\rm F})$ in thickness, while in NN mode (at $T>T_{c}$) - to the total resistance of a superconductor in
contact in the normal state $R_{{\rm S}\ (N)}$. Then the value and behavior of the magnetoresistance appear on a
different scale, such as specified by right axis scale in Fig. 5. When converted, the following parameters were
used: $\rho (H = 0)_{L({\rm F})} \sim 10^{-4} \Omega \cdot$cm; $\lambda_{T} \simeq 0.2 \mu$; $\mathcal {D}
\simeq 20 \mu$ (diameter of the contact); $L({\rm F})_{4{\rm K,\ 1 mA}} \simeq \lambda_{T}; L({\rm F})_{4{\rm K,\
100 mA}} \simeq \lambda_{T} (1 + ln100); L_{{\rm S} (N), 78 {\rm K}} \approx 3$ mm.

From a comparison of the effects at this scale with an estimated above weak-localization correction (1) it
follows that the latter is really too small to cause negative magnetoresistance $\delta \rho_{H} / \rho (H = 0)$
described by curve 1 in Fig. 5 at a current of 1 mA. At the same time, at 78 K (curve 2), in the absence of NS
boundary in contact, the negative value of $\delta \rho_{H} / \rho (H = 0)$ in the field of $10^{3}$ Oe is
comparable with the magnitude of the reduction in the resistance which accompanies the degradation of
weak-localization resistive contribution of interference nature in a magnetic field. The traces of that
contribution can be presumably seen at the indicated temperature, regardless, in general, from the value of
transport current (at least up to 100 mA).

Thus, the most likely cause of the negative magnetoresistance observed in NS mode for a heterojunction at
$T<T_{c}$ and $I=1$ mA (curve 1) is the reduction, as magnetic field increases, of some positive spin-dependent
resistive contribution which originally exists in zero magnetic field at the NS boundary inside the
superconductor (see Fig. 1) under Andreev reflection. Such a contribution in the form of a negative addition to
the conductivity of NS system may, for example, occur in conditions of itinerant magnetism of conduction
electrons, which determines the spin polarization of the current converted at the NS boundary into spinless
supercurrent due to the difference in spin densities in subbands, thus imposing a ban on certain processes of
Andreev reflection (the effect of spin accumulation [16]).

Earlier in Ref. [11, 12], we have already obtained an evidence of the existence of such a contribution in the
absence of magnetic field from the results of studying temperature properties of Andreev reflection in contacts.
However, in a magnetic field, the relative magnitude of the above correction in our contacts obviously depends on
the product of two probabilities, namely, the probability to preserve the dispersion of spin subbands along the
length \emph {L} in the \emph {N} (F) area, ie, on the length of spin relaxation $\lambda_{s}$, and the
probability of conservation of Andreev reflection. The latter depends on the electron-hole coherence in a
magnetic field characterized by the spatial coherence length $\xi (H)$ where time - reversal \emph {e-h}
trajectories are superimposed [17]. (An estimation in terms of an alternative mechanism of the occurrence of
positive addition to resistance - coherent electron - hole scattering by impurities [18] - in our $N({\rm F})/S$
systems leads to the value $L_{N({\rm F})}/l_{N({\rm F})} \simeq $ 100 times smaller due to short mean free
path.)

When $\lambda_{s}>L_{N({\rm F})}$ and yet $\xi(H) \geq L_{N({\rm F})}$, the addition discussed retains its value
in the magnetic field and the magnetoresistance can not manifest itself until reaching the fields for which
$\xi(H)$ becomes less than $L_{N({\rm F})}$, as it demonstrates curve 1 in Fig. 5. From the behavior of this
curve and estimations it follows that the condition $\xi(H)/L_{N({\rm F})} \geq 1$ is violated in the fields
higher than about 2 kOe:
\begin{equation}\label{2}
\xi_{N}(H)=\sqrt{2\lambda _{\rm B}\cdot R_{\rm L}}\Rightarrow \xi_{N}(H=2\cdot10^{3}{\rm Oe})\approx 2\cdot
10^{-5}{\rm cm} \approx \lambda_{T}= L_{N({\rm F}), 1 {\rm mA}}
\end{equation}
where $\lambda_{\rm B}$ and $R_{\rm L}$ are the de Broglie wavelength and the Larmor radius, respectively, for
LaO(F)FeAs the Fermi velocity of which is $v_{\rm F} \approx 10^{7}$ cm/s [19]. Thus, in fields higher than 2 kOe
the behavior of magnetoresistance in contact with N ({\rm F}) area, about 0.2 $\mu$ in thickness, at 1 mA and
$T<T_{c}$ could qualitatively obey the following pattern:
\begin{eqnarray}\label{3}
    \frac{\Delta\rho_{H}}{\rho_{L_{N(\rm F)}}}&=&\frac{\lambda_{s}}{L_{N(\rm F)}}\cdot \frac{P^{2}}{1 -
    P^{2}}\left[\frac{\xi_{N}(H)}{L_{N(\rm F)}} - 1\right];  \qquad (H\geq 2\cdot 10^{3}{\rm Oe})  \\
    P& = &(\sigma_{\uparrow} - \sigma_{\downarrow})/\sigma ;~ \sigma = \sigma_{\uparrow} +
  \sigma_{\downarrow};\nonumber
\end{eqnarray}
where $\sigma, \sigma_{\uparrow}, \mbox{and}\ \sigma_{\downarrow}$ are the total and spin-dependent
conductivities; \emph {P} is the spin polarization. It is assumed that all the current is converted into a
supercurrent with Andreev reflection probability equal to 1. It is provided in our heterojunctions due to an
ideal (barrier-free) NS boundary and the regime $eV(l_{el} / L_{N (\rm F)}) \sim 10^{-6} \ll \Delta_{\rm
LaO(F)FeAs}$ at the current of 1 mA. The curve corresponding to Eq. (3) with using Eq. (2) and the previously
estimated parameters $P \sim 60 \%$ [16] and $\lambda_{s} \simeq L_{N ({\rm F})} \approx 2 \cdot 10^{-5}$ cm is
shown in Fig. 5 (dotted curve 4) together with the experimental curve in the same scale of the right axis.

At a current 100 mA and $T = 4{\rm K} \ll T_{c}$, the voltage $V^{*} = V(l_{el} / L_{N (\rm F)})$ which
determines the magnitude of the jump of the distribution function $eV^{*} (- \partial f_{0} / \partial E)$, and
the gap $\Delta$ is very close in magnitude, so, the contribution from Andreev reflection (spin accumulation)
becomes negligible, although, generally speaking, the NS mode still persists together with spin polarization of
transport current in N(F) region. Close to this situation apparently takes place in the contacts Cu/FeSe, but at
a current 1 mA, as evidenced by the positive magnetoresistance (Fig. 2).

Thus, summarizing the analysis, one can conclude that the negative magnetoresistance represented by curve 1 in
Fig. 5 is at least two orders of magnitude greater than the expected value of the weak-localization correction.
It is possible only if the probability of Andreev reflection is close to unity, which is realized only in the
mesoscopic heterojunction Cu/LaO(F)FeAs in the regime $(T + eV^{*}) \ll \Delta$. In the regime $(T + eV^{*})
\gtrsim \Delta$, a negative contribution of the same value to the magnetoresistance should be absent, as is
observed (see Fig. 2 for Cu/FeSe and curves 2 and 3 in Fig. 4 for Cu/LaO(F)FeAs), due to possible degradation of
Andreev reflection as a necessary condition for spin accumulation at NS boundary.

This circumstance, as well as the hysteresis of the magnetoresistance observed in all the above energy modes,
gives reason to believe that the conductivity of superconductors studied in their ground normal state (in the
areas N (F), Fig. 1) is spin-dependent and is caused by the presence of long-range magnetic order and itinerant
magnetism of delocalized conduction electrons due, apparently, to \emph {s-d} exchange interaction of iron atoms.
Since the investigated systems are very critical to the conditions of emerging superconductivity (eg, as a
consequence of the Stoner ferromagnetic instability leading to spin fluctuations [20]), we can assume that
arising anisotropic magnetic structure of the local magnetic moments is metastable and tends to have
metamagnetism in low magnetic fields. According to the spin-dependent part of the Heisenberg Hamiltonian, the
potential of spin interaction can be written as
\begin{equation}\label{4}
    U_{m} = -(J_{m}/n_{m}) \sum_{i}\mathbf{s}_{e}\mathbf{S_{i}}f(\mathbf{r} - \mathbf{R}_{i}),
\end{equation}
where $J_{m}$ are anisotropic exchange interaction coefficients responsible for processes of ferromagnetic
($J_{1}$) and antiferromagnetic ($J_{2}$) exchange between the spins of magnetic ions $\mathbf {S}$ and those of
conduction electrons $\mathbf {\sigma}$, while $n_{m} = N_{m}/V$ (here, $N_{m}$ is the number of ions in the
stripes with the \emph {m}-exchange interaction  (see Fig. 6 of [21]). Under these conditions, total amplitude of
the scattering of electron spins by ion spins could not remain constant when both the direction and magnitude of
even minor magnetic fields change.

\begin{figure}[htb]
  \begin{center}
  \includegraphics[width=10cm]{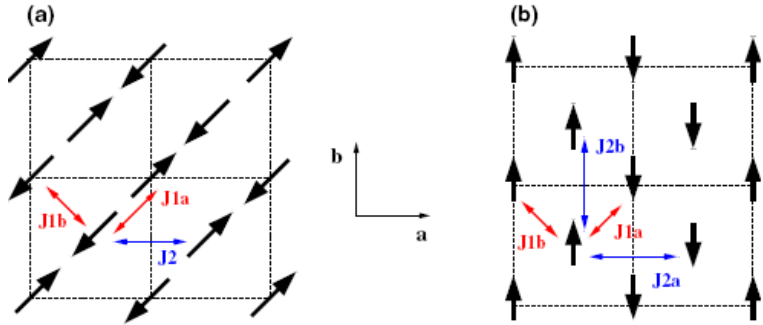}\\
  \caption{Spin magnetic structure of iron-based superconductors (parent): (a)
oxyarsenides and (b)  chalcogenides. Exchange rates J1a, J1b, J2a, and J2b refer to AFM, FM, AFM, and FM
orderings of  spins of iron ions, respectively [20, 22].}\label{6}
\end{center}
\end{figure}

Indeed, given the nature of the percolation of current flow along the stripes with different spin ordering,
spin-dependent part of the conductivity is reasonably represented in the form

\begin{equation}\label{5}
\overline{\sigma_{spin}} = \sigma_{J_{1}} + \sigma_{J_{2}},
\end{equation}
and corresponding resistance as

\begin{equation}\label{6}
\overline{\rho_{spin}} = \rho_{J_{1}}\rho_{J_{2}}/(\rho_{J_{1}} + \rho_{J_{2}}).
\end{equation}
As known [14, 23], everything else being equal, $\rho_{J_{2}}\gg\rho_{J_{1}}$ since the AFM scattering
($J_{2}<0$) admits a spin-flip processes thereby increasing the scattering amplitude which allows to consider the
component $\rho_{J_{1}}$ corresponding to the FM exchange prevailing. Since the resistance is proportional to the
square of the scattering amplitude which, in turn, according to Eq. (4), is proportional to $-(J_{m}/n_{m})
(\mathbf {s}_{e} \mathbf {S})_ {s'_{e} s_{e}}$, after summing over the final spin orientations and averaging over
the initial orientations, the component $\rho_{J_{1}}$ in the first Born approximation must have the form [21]:

\begin{equation}\label{7}
 \rho_{spin1} = \frac{3\pi m}{2e^{2}\hbar\varepsilon_{\rm F}}S(S + 1)J_{1}^{2}/n_{1}
\end{equation}
Here, \emph {m} and \emph {e} are the electron mass and charge, $\varepsilon_{\rm F}$ is the Fermi energy. From
this it follows that if the magnetic field reorients a part of spins in stripes reducing, in particular, $n_
{1}$, it will lead to the following dependence of the magnetoresistance on magnetic field:

\begin{equation}\label{8}
    \frac{\Delta\overline{\rho_{spin}}}{\rho_{0}} \sim J^{2}_{1}(n_{1})^{-1}\sim f_{ev}(H),
    \end{equation}
where, according to the experiment, $f_{ev}(H)$ is an even function of the magnetic field.

\begin{figure}[htb]
  \begin{center}
  \includegraphics[width=10cm]{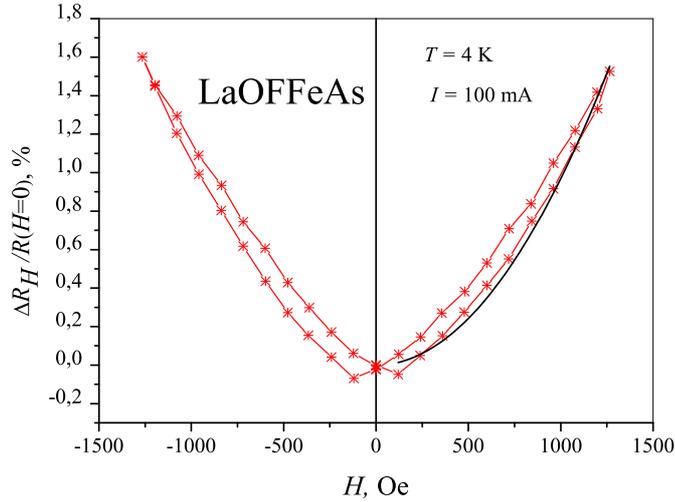}\\
  \caption{Magnetoresistance with hysteresis of
the heterojunction Cu/LaO(F)FeAs, normalized to the total contact resistance: $\ast$ represent experiment $((T +
eV^{*}) \sim \Delta$); solid curve is spin-dependent magnetoresistance in accordance with the Eq. (5).} \label{3}
\end{center}
\end{figure}

Solid lines in Figs. 2 and 7 show the magnetoresistance corresponding to Eq. (7) at $(n_{1})^{-1} \sim H^{2}$.
Note that the result obtained in the first Born approximation predicts only positive magnetoresistance,
hysteresis and different values of $\Delta \rho_{H} / \rho_{0}$ in FeSe and LaO(F)FeAs in the normal state in
regimes $(T + eV^{*}) \gtrsim \Delta$ [21].

In conclusion, we investigated non-ballistic NS heterocontacts with barrier-free boundary inside iron-based
superconductors between mesoscopic non-superconducting region and the superconducting part. We studied the
behavior of the magnetoresistance of the contacts in small external magnetic fields depending on the energy
experimental conditions, namely, temperature and bias voltage on the contact, which determine the energy of the
charge carriers, as well as on the magnetic field of the transport current, which determines the initial size of
the non-superconducting region of superconductors. It has been found that when the probability of Andreev
reflection, $r_{\rm A}$, at the NS boundary is close to unity, the conductivity increases since the fields of the
order of $10^{3}$ Oe, causing a negative magnetoresistance which is more than an order of magnitude greater than
the magnitude of the possible contribution of weak-localization nature.

The increase in conductivity, as well as the hysteresis effect which cannot be eliminated even in the mode with
the probability of Andreev reflection at the NS boundary close to zero, indicate the spin-dependent nature of
transport in the normal ground state of chalcogenides and arsenide oxypnictides. In condition $r_{\rm A} \approx
1$, that nature of transport is testified by the presence of another addition to the resistance of normal area in
zero magnetic field. Its value corresponds to spin accumulation at dispersion of spin subbands with a
polarization factor, more than 50 \% in magnitude. The addition should, as observed, decrease under the
destruction of \emph {e-h} coherence in the Andreev reflection when applying a magnetic field. Spin-dependent
character of transport also manifests itself in the behavior and the value of magnetoresistance in conditions
$r_{\rm A} \approxeq 0$. In particular, positivity and evenness (with respect to the magnetic field
direction) of the magnetoresistance in this transport mode indicate the possibility of Stoner ferromagnetic
instability, at which the amplitude of spin scattering may depend on the number of ferromagnetically oriented
local moments in stripes with ferromagnetic interaction of local spins with spins of itinerant electrons.

Research was supported in part by a grant from the Russian Foundation for Basic Research "RFBR
14-02-92002-NNS-a."

\end{document}